\documentclass[prl,aps, babel,twocolumn]{revtex4}

\usepackage{amsmath,amssymb,amstext,amscd,amsthm,amsopn,
graphicx,indentfirst,mathrsfs}

\begin{document}

\title{Inside an evaporating two-dimensional charged black hole}
\author{Amos Ori}
 \affiliation{Department of Physics,\\
 Technion---Israel Institute of Technology, Haifa, 32000, Israel}

\begin{abstract}
We investigate the inner structure of an evaporating charged black hole, 
within the context of semiclassical dilaton gravity in two dimensions.
The matter fields are charged, allowing the evaporation of both the
 mass and charge of the black hole. We find that the semiclassical 
effects cause the inner horizon to expand (by a finite factor), 
rather than to shrink to a point singularity. 
Although this expansion is a quantum phenomenon, the overall expansion factor
is found to be independent of the magnitude of the quantum terms in the effective theory.
\end{abstract}
\maketitle

The internal structure of charged black holes (BHs) has been the subject of
many investigations over several decades. The Reissner-Nordstrom solution
admits a perfectly regular inner horizon (IH) inside the BH, as opposed to
the spacelike singularity of the uncharged Schwarzschild solution. Classical
perturbations are known to develope instabilities on the IH, which lead to a
null curvature singularity \cite{pi}. However, this singularity is found to
be weak and non-destructive, in the sense that its tidal effect on a
physical object is bounded and small \cite{ori91}. A similar situation was
observed in a spinning BH \cite{ori00}.

When semiclassical effects are taken into account, the black hole evaporates. In the
case of a charged BH there are two distinct semiclassical phenomena: The
electromagnetic Schwinger effect, and the gravitational Hawking effect. The
evaporation of a charged BH involves both phenomena. An important question
is, how does this evaporation affect the internal geometry. The goal of the
present paper is to address this problem within a simplified two-dimensional (2d)
model.

Callan et al \cite{CGHS} developed a dilatonic 2d model of an
uncharged BH. This model as well as similar ones were subsequently
investigated by many authors \cite{russo}. Similar models of a charged BH
(yet with uncharged matter fields) were later developed \cite{MNY} \cite
{Elizalde}. Nojiri and Oda \cite{NO} then considered a dilatonic model with
a charged quantum matter field. In their model, however, the
dilaton-electromagnetic coupling took a non-standard form, which led to
undesired causal structure of the charged BHs already at the classical
level. This problem was later fixed in Ref. \cite{ori}.

The model \cite{ori} consists of gravity in two dimensions, coupled to a
dilaton $\phi $, electromagnetic field $F_{uv}$, and $N$ charged matter
fields. These fields were taken to be massless chiral fermions, following
Nojiri and Oda \cite{NO}. The dilaton-electromagnetic coupling was taken
as $e^{-2\phi }$ (rather than the coupling $e^{2\phi }$ used in Ref. \cite
{NO}). The analysis in \cite{ori} focused primarily on the evolution of the
various fields outside the evaporating BH. In particular the evaporation
rate of the BH's mass and charged was determined. In this paper we shall use
the same model to explore the evolution of geometry \textit{inside} the BH,
with special emphasize on the dynamics near the IH.

Recently Frolov, Kristjansson, and Thorlacius (FKT) \cite{fkt} developed a
similar model of 2d dilaton gravity coupled to a charged matter field. The
main difference is that in Ref. \cite{fkt} the matter field is 
bosonized before semiclassical corrections are calculated. 
Also this field may be either massive or massless. In the
massless case the field equations are closely related to those of Ref. \cite
{ori}. FKT evolved the field equations (for a massless field) numerically in
order to explore the dynamics inside the evaporating charged BH. They
concluded that the combination of the Schwinger and Hawking effects turns 
the IH into a spacelike singularity where $\phi $ diverges to 
$+\infty $. 
We employ here the usual translation of the 2d dilatonic terminology to that
of a four-dimensional (4d) spherical model, in which 
the area coordinate is mapped to $e^{-2\phi }$.
Then in this language, the inner horizon was found in \cite{fkt} to shrink
to a point singularity.

Previously several authors studied numerically the backreaction of
Schwinger's pair-production effect on the internal structure of a charged
BH, in either 4d \cite{ns} \cite{piran} or 2d gravity \cite{fkt0}. These
analyses, however, did not incorporate the Hawking effect and hence the BH did
not evaporate, a fact which drastically changes the global structure. Also
Balbinot and Brady \cite{bb} studied the backreaction to the Hawking effect
inside a 2d charged BH, but the Schwinger effect was not included, and the
BH was a non-evaporating one. To the best of our knowledge Ref. \cite{fkt}
is the only work published so far which analyzes the inner structure of an
evaporating semiclassical charged BH.

In this paper we explore the field equations of Ref. \cite{ori}
analytically, and reveal the internal structure of a semiclassical charged
BH which evaporates its both mass and charge. We find a behavior rather
different than that reported in \cite{fkt}. On approaching the IH, the
semiclassical effects cause the area coordinate to \textit{expand} (to a
finite value), rather than to shrink to a point singularity. This
semiclassical expansion is linear in the Eddington coordinates. These
analytical results were also confirmed by numerical simulations of our field
equations. The origin of the discrepancy between FKT's analysis and our
results still needs be understood. In this paper we briefly present our analysis and
describe the main results. A more detailed description will be given in a
separate paper \cite{f}.

The classical action and the semiclassical corrections are given in Ref. 
\cite{ori} (it is the same as in Ref. \cite{NO}, apart from the modified
dilaton-electromagnetic coupling). Our
notation here follows that of Ref. \cite{ori}, except that we choose our
units such that the cosmological constant $\lambda =1$. We express the
metric in double-null coordinates, $ds^{2}=-e^{2\rho }dudv$. (Both $u$ and $%
v $ are future-directed, as usual, and the lines of constant $u$ or constant 
$v $ are respectively parallel to past or future null infinity.) The
electromagnetic field is expressed as $F_{uv}=(g/\sqrt{6})e^{2(\phi +\rho
)}q $, where $g>0$ denotes the electromagnetic coupling constant. The system
thus consists of three unknowns, $\phi ,\rho ,q$. We then transform to new
variables $R\equiv e^{-2\phi }$ and $S\equiv 2(\rho -\phi )$. The resulting
field equations \cite{eqs} include a set of three hyperbolic equations for
the three unknowns $R,S,q$: 
\begin{eqnarray}
R_{,uv} &=&({q}^{2}/R^{2}-1)e^{S}-K\rho _{,uv}\ ,  \label{Ruv} \\
S_{,uv} &=&-2({q}^{2}/R^{3})e^{S}+K\rho _{,uv}/R\,,  \label{Suv} \\
q_{,uv} &=&-Kg^{2}(q/R^{2})e^{S}\,,  \label{quv}
\end{eqnarray}
where $K\equiv N/24$ (representing the magnitude of the semiclassical
effects), and $\rho =(S-\ln R)/2$. In addition there
are two constraint equations, which both take the form 
\begin{equation}
R_{,ww}-R_{,w}S_{,w}+\hat{T}_{ww}=0~,  \label{Rww}
\end{equation}
where hereafter $w$ stands for either $u$ or $v$, and 
\begin{equation}
\hat{T}_{ww}=K[(\rho _{,ww}-\rho _{,w}^{\,2})+(Kg)^{-2}q_{,w}^{2}\,+\hat{z}%
_{w}(w)]\,
\end{equation}
represents semiclassical fluxes of energy in the $u$ or $v$ directions. The
functions $\hat{z}_{u}(u)$ and $\hat{z}_{v}(v)$ are determined from
the boundary conditions.

The classical equations are obtained by formally setting $K=0$. Then $\hat{T}%
_{ww}$ vanishes, and $q$ is set to a constant. The general
solution then forms a two-parameter set, parametrized by the mass $m$ and
the charge $q$, to which we shall refer here as the two-dimensional
Reissner-Nordstrom (2dRN) solution. This solution is most easily expressed
in diagonal coordinates \cite{MNY}. For the present analysis,
however, we shall need the classical solution in the double-null coordinates 
$u,v$.

A key function which appears in the classical solution is $H(R)\equiv
R-2m+q^{2}/R$. We are considering here the non-extreme case, $m>q>0$. Then
this function has two zeros, at $R_{\pm }=m\pm (m^{2}-q^{2})^{1/2}$.
Correspondingly the charged BH admits two horizons: an event horizon (EH) at 
$R=R_{+}$ and an IH at $R=R_{-}$. The latter also functions as a Cauchy
horizon (CH) for initial hypersurfaces outside the BH.

For the analysis below we shall need the classical solution in between the
two horizons, namely at $R_{-}<R<R_{+}$, in the ''Eddington-like'' gauge in
which both $R$ and $S$ depend on the single variable $x\equiv u+v$. The
solution consists of the two functions $R(x)$ and $S(x)$, which satisfy 
\begin{equation}
\ e^{S}=-H(R)\,,\,\,R_{,x}=H(R)\,.  \label{classic}
\end{equation}
$R(x)$ is formally expressed through its inverse function, 
\[
x(R)=\int_{m}^{R}[1/H(R^{\prime })]dR^{\prime }\,.
\]
(The lower integration limit was set here to $m$ for concreteness.) Note
that $H$ is negative at $R_{-}<R<R_{+}$, and correspondingly $x$ tends to $-%
\infty $ at the EH and to $+\infty $ at the IH. 
In a classical
collapse scenario the EH (IH) is located at $u\rightarrow -\infty $
($v\rightarrow +\infty $) and finite $v$ ($u$). 

We also define the two parameters 
\[
\kappa _{\pm }\equiv \frac{1}{2}(dH/dR)_{R=R_{\pm }}=\pm %
[(m^{2}-q^{2})^{1/2}/R_{\pm }]\,.
\]
Note that in our notation $\kappa _{+}>0$ and $\kappa _{-}<0$, 
and that $\kappa _{\pm}$ depend solely on $q/m$.

The domain $R_{-}<R<R_{+}$ is naturally divided into three regions: (i) The 
\textit{near-EH zone}, namely the region of very negative $x$ ($-x>>1$), (ii) The 
\textit{near-IH zone}, $x>>1$, and 
(iii) The \textit{central zone}, namely the region of moderate $x$ values.
In both the near-EH and near-IH zones $e^{S}$
is exponentially suppressed, like $e^{2\kappa _{+}x}$ and $e^{2\kappa _{-}x}$
respectively. 

Consider next the subset of classical solutions with fixed ratio $%
q/m=\alpha $. It depends on a single parameter $m$. One can easily verify that
all solutions in this one-parameter family are related by a simple scaling
rule: 
\[
R(x;m)=mR_{0}(x)\,,\,\,S(x;m)=\ln m+S_{0}(x)\,, 
\]
where $R_{0}(x)\equiv $ $R(x;m=1)$ and $S_{0}(x)\equiv $ $S(x;m=1)$ are the
functions corresponding to unit mass (and $q=\alpha $).

As long as the BH is macroscopic ($m>>K$), the semiclassical effects outside
the BH are weak in a local sense; Namely, in the neighborhood of each point
the metric is that of a classical 2dRN solution weakly perturbed by the
semiclassical fluxes. These weak fluxes accumulate over long time scales,
however, and lead to slow decay of $m$ and $q$. We denote the
slowly-drifting values of $m$ and $q$, 
measured along the EH, by $m_{eh}(v)$ and $q_{eh}(v)$
respectively. In Ref. \cite{ori} we calculated the rate of change of these
quantities, and analyzed their long-term evolution. It was shown that for $%
g<1/2$ the ratio $q/m$ approaches a constant 
\begin{equation}
\alpha =\sqrt{1-2g}\,/(1-g)\,\,.  \label{smallg}
\end{equation}
Furthermore, if $q/m$ is initially $\alpha $, then this ratio is preserved
during the evaporation. In this case the mass and charge evaporate at
constant rates (apart from corrections of order $K/m_{eh}$ \cite{log}). 
Setting $v=0$
at the moment when (by extrapolation) $m_{eh}$ vanishes, we may write 
\begin{equation}
m_{eh}(v)=-\beta v\,,\,\,\,q_{eh}(v)=-\alpha \beta v\,,  \label{rates}
\end{equation}
where $\beta =Kg^{2}m/R_{+}$.

In what follows we shall consider the case $g<1/2$. For simplicity we shall
also assume that the BH's initial $q/m$ is the preferred value (\ref{smallg}) 
(but our main conclusions do not crucially depend on this assumption \cite
{f}).

We turn now to discuss the evolution inside the BH, assuming that the latter
is macroscopic (throughout the rest of this paper by ''macroscopic'' we mean 
$R_{-}>>K$, which also implies $m>>K$). In fact the above statement about
the slow semiclassical evolution also holds inside the BH at $R>R_{-}$
(except at the very neighborhood of $R_{-}$, as we discuss below). We can
then express the solution for $R,S,q$ in a perturbative manner, where the
leading order is the classical 2dRN solution with slowly-varying parameters $m$ 
and $q=\alpha m$. We shall refer to this approximation as the \textit{adiabatic
approximation}. In the range $R_{-}<R<R_{+}$ which concerns us here, this
expansion takes the form 
\begin{eqnarray}
R(u,v) &=&m_{eh}[R_{0}(u+v)+\varepsilon R_{1}(u,v)+O(\varepsilon^{2})]\,\,,
\label{RSqad} \\
S(u,v) &=&\ln (m_{eh})+S_{0}(u+v)+\varepsilon S_{1}(u,v)+O(\varepsilon ^{2})\,\,, 
\nonumber \\
q(u,v) &=&m_{eh}[\alpha +\varepsilon q_{1}(u,v)+O(\varepsilon ^{2})]\,\,, 
\nonumber
\end{eqnarray}
where $\varepsilon \equiv K/m_{eh}$ and, recall, $m_{eh}=-\beta v$. The field
equations (\ref{Ruv}-\ref{Rww}) are then automatically satisfied at the
leading order $\varepsilon^{0}$. At the next order $\varepsilon^{1}$ one obtains a
set of linear field equations for the three new functions $F_{1}(u,v)$,
where hereafter $F$ will stand for the three unknowns $R,S,q$. The initial
data for $F_{1}$ at the EH are discussed in Ref. \cite{f}.

For the present analysis we do not really need the specific solution for the
perturbation fields $F_{1}$. All that we need is to make sure that these
fields are well controlled inside the BH (and hence the perturbation term $%
\propto \varepsilon $ is negligible compared to the leading classical term). A
close look at the field equations for $F_{1}$ reveals \cite{f} that indeed
these fields are well controlled 
in the near-EH and central zones; 
However, in the near-IH zone (namely at large positive $u+v$) 
$R_{1}$ and $q_{1}$ evolve linearly in $u$ and $v$ (see below) and get
large values. This linear drift formally invalidates the above perturbative
scheme at large positive $u+v$. Since in this paper we are primarily
interested in the behavior near the IH, we shall develope now another
perturbative scheme which will be valid at large positive $u+v$ despite the
linear drift in $R$ and $q$.

We first bring Eqs. (\ref{Ruv}-\ref{quv}) to a standard
hyperbolic form by isolating 
$R_{,uv}$ and $S_{,uv}$. We obtain a system of the form 
\begin{equation}
F_{,uv}=A_{F}(R,q)e^{S}+KB_{F}(R)R_{,u}R_{,v}\,,  \label{Fhyp}
\end{equation}
with $A_{q}=-Kg^{2}q/R^{2}$ and $B_{q}=0$. For $F=R,S$ the coefficients $%
A_{F},B_{F}$ are certain functions of their arguments (and $K$) which we
write explicitly in Ref. \cite{f}. For the present analysis we shall not
need the $B$-functions, and also it will be sufficient to express 
$A_{R}$ and $A_{S}$ at the leading order $K^{0}$: 
\[
A_{R}\simeq {q}^{2}/R^{2}-1\,,\,\,\,A_{S}\simeq -2{q}^{2}/R^{3}\,. 
\]
In Ref. \cite{f} we analyze the effect of the terms proportional to 
$R_{,u}R_{,v}$ on the near-IH asymptotic behavior and find it to be
negligible 
(note that $R_{,w}$ vanishes at the IH at the leading order). 
Omitting these terms, the evolution equations
take the simple form 
\begin{equation}
F_{,uv}=A_{F}(R,q)e^{S}\,.  \label{Fuv}
\end{equation}

Our goal is to explore the asymptotic behavior of this system at large $u+v$. 
To this end we first consider the classical case (namely $K=0$ and $%
q=const $). From Eq. (\ref{classic}) it follows that $S$ decreases at large $%
x$ linearly, $S\approx 2\kappa _{-}(u+v)$. Consequently $e^{S}$ becomes
negligible, and Eq. (\ref{Fuv}) reduces to 
\begin{equation}
F_{,uv}\cong 0\,.  \label{Fuv0}
\end{equation}
The above asymptotic form of $S$ obviously satisfies this equation. The
equation for $R$ implies $R\cong R_{u}(u)+R_{v}(v)+const$, and it is indeed
satisfied with $R_{u}=R_{v}=0$ and $const=R_{-}$.

Turn now again to the system (\ref{Fuv}) in the semiclassical case. The most
crucial ingredient in the above classical asymptotic behavior is the global
decaying factor $e^{S}$. There is no reason to expect that this behavior
will be sensitive to small modifications in the coefficients $A_{F}$. Thus,
the system (\ref{Fuv}) seems to admit a rather generic class of asymptotic
solutions at large $u+v$, in which $S$ decreases linearly (or alike) in $u$
and $v$, leading to Eq. (\ref{Fuv0}). Then the asymptotic solution should
take the simple form 
\begin{equation}
F\cong F_{u}(u)+F_{v}(v)\,  \label{Fuv0s}
\end{equation}
(up to exponentially-small corrections). The claim that our system of
evaporating BH indeed falls into this class of asymptotic solutions will be
justified below, by determining the actual functions $F_{u}$ and $F_{v}$.

The functions $F_{u},F_{v}$ must satisfy certain consistency conditions in
order to enable this simple asymptotic behavior. First, $S_{u}$ and $S_{v}$
must decrease sufficiently fast in $u$ and $v$ respectively (a linear
decrease, as in the classical solutions, is sufficient). Second, since the
functions $A_{F}$ diverge at small $R$ (in fact  $A_{F}$ has poles at $R=0$
and $R=K$), the functions $R_{u}$ and $R_{v}$ must have a form which
guarantees that $R$ will increase (or at least not decrease) with $u$ and $v$.

In fact, the analysis below indicates that the asymptotic solution (\ref
{Fuv0s}) takes the more specific form 
\begin{equation}
F\simeq a_{F}u+b_{F}v+c_{F}\,,  \label{Fabc}
\end{equation}
where $a_{F},b_{F},c_{F}$ are certain constants. In terms of these
parameters, the above consistency conditions reduce to 
(i) $a_{S},b_{S}<0 $, and (ii) $a_{R},b_{R}\ge 0$. The sign of 
$a_{q} $ and $b_{q}$ is unimportant. Note that with such a linear behavior,
at large $u+v$ the source terms in Eqs. (\ref{Fuv}) are all bounded by $%
(u+v)^{2}e^{(a_{S}u+b_{S}v)}$, which indeed has a negligible effect and
justifies the approximation (\ref{Fuv0},\ref{Fuv0s}).

The determination of the required functions $F_{w}(w)$ is done by matching
the near-IH approximation (\ref{Fuv0}) to the adiabatic expansion (\ref
{RSqad}). This matching is done in the {\it intermediate zone} 
\begin{equation}
1<<u+v<<R_{-}/K\,.
\end{equation}
The inequality $u+v>>1$ guarantees that $e^{S}\propto e^{2\kappa
_{-}(u+v)}<<1$ and the near-IH approximation (\ref{Fuv0}) is valid. The
inequality $u+v<<R_{-}/K$ means that in the adiabatic expansion (\ref{RSqad}%
) the terms $F_{1}$ (which are $\propto K$ but undergo linear drift at large 
$u+v$) are unimportant, and the adiabatic approximation holds. A consistency
condition for this matching is that in the intermediate zone $F_{,u}$ must
be independent of $v$, and vice versa. If in addition $F_{,w}$ is
independent of $w$, then the linear form (\ref{Fabc}) holds as well.

The parameters $a_{S}$ and $b_{S}$ are nonvanishing already at the classical
level ($a_{S}=b_{S}= 2\kappa _{-}$). The semiclassical effects lead to 
$O(\varepsilon )$ corrections to these parameters, but these small corrections are not 
needed here. On the other
hand, $a_{R},b_{R},a_{q},b_{q}$ all vanish at the classical level, and our
goal is to calculate these four parameters at order $\varepsilon ^{1}$. It
will be sufficient to calculate $c_{F}$ at order $\varepsilon ^{0}$, because
the linear drift in $F$ is not sensitive to these constants.

Consider first the semiclassical evolution of $q$. Integrating the field
equation (\ref{Fuv}) for $F=q$ with respect to $u$ or $v$, yields 
\[
q_{,w}\cong q_{,w}^{(eh)}-Kg^{2}\int\nolimits_{-\infty }^{\infty %
}(qe^{S}/R^{2})\,\,d\hat{w}\,,
\]
where $\hat{w}$ is the null coordinate other than $w$ (e.g. $\hat{w}=v$ when 
$w$ stands for $u$), and the integration is done along a line of constant $w$.  
Since we only need $q_{,w}$ at order $K^{1}$, it is sufficient to
evaluate the integrand at the leading order $\varepsilon ^{0}$, namely it can
be taken as  $\alpha e^{S_{0}}/R_{0}^{2}$. Taking the upper integration limit
as $\infty $ is justified because of the exponential decay of $e^{S}$ at
large $u+v$. The integration is then straightforward and the integral
becomes $q/R_{-}-q/R_{+}$. [Note that unlike $q$ and $R_{\pm }$, $q/R_{\pm }$
is independent of $m_{eh}$ and only depends on $\alpha $.] Noting that $q_{,u}^{(eh)}=0$ (e.g. from
regularity) and $q_{,v}^{(eh)}\equiv q_{eh,v}$ is $-\alpha \beta
=-Kg^{2}q/R_{+}$, we obtain at the intermediate zone 
\[
q_{,u}\cong Kg^{2}(q/R_{+}-q/R_{-}) \equiv a_{q} ,\,
q_{,v}\cong -Kg^{2}q/R_{-} \equiv b_{q} .
\]
Both quantities are independent of $v$ and $u$, confirming the
 linear form (\ref{Fabc}) for $q$.

Next consider the semiclassical evolution of $R$. Here again we evaluate 
$R_{,u}$ and $R_{,v}$ at the intermediate zone. To this end we use the
constraint equation (\ref{Rww}). 
Since we only need $R_{,w}$ at order $K^{1}$, 
it will be sufficient to calculate 
$\hat{T}_{ww}$ at its leading order---namely
that of a test charged field in a classical 2dRN background. 
This test-field $\hat{T}_{ww}$ 
is to be evaluated at the near-IH zone of this
2dRN geometry, where $\rho _{,ww}\cong 0$ and $\rho
_{,w}\cong \kappa _{-}$. Also the boundary conditions 
imply \cite{ori} $\hat{z}_{u}=\kappa _{+}^{2}$
and $\,\hat{z}_{v}=0$. One then finds that $\hat{T}_{uu}$ and $%
\hat{T}_{vv}$ are just constants. 
Equation (\ref{Rww}) then yields $R_{,w}\cong \hat{T}_{ww}/S_{,w}$ 
(the corresponding homogeneous solution $R_{,w}=e^{S}$ may be 
ignored at large $u+v$). Substituting $S_{,w}\cong 2\kappa _{-}$ we find 
\[
R_{,u}\cong K[g^{2}(q/R_{-}-q/R_{+})^{2}\,+(\kappa _{+}^{2}-\kappa
_{-}^{2})]\,/2\kappa _{\_}\equiv a_{R}\,\,
\]
and 
\[
R_{,v}\cong K[(gq/R_{-})^{2}\,-\kappa _{-}^{2}]\,/2\kappa _{-}\equiv b_{R}\,.
\]
Again, both $R_{,u}$ and $R_{,v}$ are independent of $v$ and $u$, which
justifies the linear form (\ref{Fabc}).

As was discussed above, our near-IH approximation will only be valid if both 
$a_{R}$ and $b_{R}$ are positive. Noting that Eq. (\ref{smallg}) implies $%
\kappa _{+}=g$, one obtains 
\[
a_{R}=Kg^{2}(R_{+}/R_{-}+R_{-}/R_{+})\,\,,\,\,b_{R}=Kg^{2}(R_{+}/R_{-})\,. 
\]
Both parameters are positive as desired.

The constants $c_{R}$ and $c_{q}$ are calculated from Eq. (\ref{Fabc}) by
setting $R\simeq R_{-}^{eh}$ and $q\simeq q_{eh}$ at the intermediate zone, 
where $R_{\pm}^{eh}$ denotes the value $R_{\pm}$ corresponding to 
$m_{eh}(v)$. One obtains \cite{ori} $c_{q}=c_{R}=0$.

The global picture of the region between the two horizons as emerges from
our analysis is fairly simple. Since we have set $v=0$ at the
end-point of evaporation, the CH of the evaporating BH is the line $v=0$.
The EH is located at $u=-\infty$, just as in the non-evaporating case. 
$v$ is negative and increasing along the EH, and $u$ is positive 
and increasing along the CH (starting from $u\approx 0$).
In both the near-EH and the central zones the spacetime is well described 
by the adiabatic 
solution (\ref{RSqad}), in which the mass and charge slowly evolve with $v$.
In particular, $R$ (like $q$) decreases linearly along the EH. In the
near-IH zone all dynamical variables drift linearly in $u$ and $v$. Along
the CH $R$ (like $-q$) grows linearly, $R\simeq a_{R}u$, 
starting from $R\simeq 0$ 
at $u\simeq 0$ and subsequently growing to macroscopic values.

In considering the near-IH behavior of $q$, one recognizes that both $a_{q}$
and $b_{q}$ are negative. In fact the charge flips its sign before the CH
is approached, 
and along the CH $q\simeq a_{q}u<0$. 

The positivity of $a_{R}$ and $b_{R}$ means that the effect of the
semiclassical terms near the IH is to expand the latter (namely to
increase $R$), rather than to shrink it to a point singularity $R=0$.
Consider an infalling observer which crosses the EH at some $v=v_{0}$ in the
macroscopic domain (namely $-v_{0}>>1$) and takes a typical orbit which in the
locally-2dRN background would hit the IH at some finite $u$. The
observer approaches the central zone at $v\approx v_{0}$ and 
$u\approx u_{0}\equiv -v_{0}$. 
Subsequently on heading towards the IH $v$ increases remarkably but $u$ is
essentially frozen at $u\approx u_{0}$. In the first part of the journey,
during the adiabatic phase, $R$ steadily decreases from $R_{+}$ to $R_{-}$ 
(corresponding to the entrance mass $m_{eh}(v_{0})$) 
just as in a 2dRN spacetime. 
But subsequently, on approaching the very neighborhood of
the CH, $R$ will start increase according to 
\[
R\simeq a_{R}|v_{0}|+b_{R}v\simeq R_{-}^{eh}(v_{0})+b_{R}(v-v_{0})\,.
\]
Note that this phase of increasing $R$ lasts an enormously
short proper time, due to the huge blue-shift factor near the IH. The final
value at the CH is $R\simeq a_{R}|v_{0}|$. A simple calculation shows that
this value is even greater than $R_{+}^{eh}(v_{0})$, by the factor $%
R_{+}/R_{-}+R_{-}/R_{+}>1$.

Note that this final $R$ value at the CH is independent of $K$;
Namely, although the near-IH increase in $R$ is a quantum phenomenon, its
magnitude only depends on the macroscopic parameters $m_{eh}$ and $q_{eh}$,
not on the amplitude of the quantum terms in the effective action. 
This has a simple explanation:
The rate of the near-IH semiclassical increase in $R$ is $\propto K$, but at
the same time, since $\beta \propto K$, for a given entrance mass $%
m_{eh}(v_{0})$ the evaporation time (from $m_{eh}(v_{0})$ to complete
evaporation) is $\propto 1/K$, hence their product is independent of $K$.
The same considerations apply to the semiclassical drift in $q$, and one
finds that the final charge at the CH is $2\kappa _{-}$ times the entrance
value $q_{eh}(v_{0})$, independent of $K$.

We also performed numerical simulations of the full nonlinear hyperbolic
system (\ref{Ruv}-\ref{quv}) [using the form (\ref{Fhyp}) with the exact
functions $A_{F},B_{F}$]. We started from initial data corresponding to
Eq. (\ref{rates}) along the EH. These simulations confirm the above analysis.
In particular, all quantities $F$ where found to evolve linearly in $u$ and $%
v$ in the near-IH zone, with $R$ increasing and $q$ decreasing. 
The numerically computed quantities $F_{,u}$ and $F_{,v}$
all agree with the above analytically calculated parameters $a_{F}$ and $%
b_{F}$ to within a few percents. More details of the numerical simulations
will be presented in Ref. \cite{f}.

The rapid change in $R$ as a 
function of the observer's proper time in the near-IH zone rases the 
question of whether the semiclassical theory is valid 
in this region. Scalars made of gradients, such as 
$e^{-2\rho} \phi_{,u} \phi_{,v}$, get enormously large (though finite) values on 
approaching the CH. Yet the effective gravitational coupling constant 
$e^{\phi}$ \cite{CGHS} remains small (except near $u=v=0$). 
Nevertheless the present analysis makes one point very clear: 
When the semiclassical theory comes to its limit of predictability 
(either at $v=0$ or at an earlier stage where gradient-based scalars become large), 
$R_{,v}$ is {\it positive} rather than negative. 
This fact may have important implications to the issue of extension 
of geometry beyond the CH, which we shall not address here. 

Finally we comment on the relevance of our results to the real 4d world. 
A key ingredient in the above 2d analysis is the negative sign 
of $\hat{T}_{uu}$ and $\hat{T}_{vv}$ of a {\it test} field near the IH.
 In the case of a 4d spherical charged BH it is possible to bring the 
 semiclassical field equations to a form very similar to the above 2d model. 
 This similarity strongly suggests that in the 4d case too, 
 the crucial factor which will determine whether the 
 CH expands or contracts will be the sign of $\hat{T}_{uu}$ and $\hat{T}_{vv}$ 
 near the IH, for a test quantum field in the Reissner-Nordstrom geometry. 
 Unfortunately in the 4d case there is no simple expression for $\hat{T}_{\mu \nu}$. 
 Nevertheless certain methods were developed for calculating 
 $\hat{T}_{\mu \nu}$ in the 
 Reissner-Nordstrom geometry outside the EH \cite{anderson}. 
 The present analysis emphasizes the importance of extending the calculation of 
 $\hat{T}_{\mu \nu}$ to the interior of the 4d Reissner-Nordstrom BH, 
 particularly to the neighborhood of the IH.

I am grateful to Larus Thorlacius for warm hospitality and fruitful
discussions. I also thank Joshua Feinberg and Joseph Avron for
helpful comments.


\begin{thebibliography}{99}
\bibitem{log}   These $O(K/m)$ terms lead to logarithmic 
corrections in Eq. (\ref{rates}). Nevertheless these
corrections are negligible in the macroscopic case and will be
ignored here.
\end{thebibliography}
\end{document}